\documentclass{osa-article}

\usepackage{color}
\usepackage{comment}
\usepackage{lineno}
\usepackage{multirow}
\usepackage[toc,page]{appendix}
\usepackage{siunitx}

\journal{osajournal}

\articletype{Research Article}

\begin{document}
\title{Anti-reflection coating with mullite and Duroid for large-diameter cryogenic sapphire and alumina optics}

\author{Kana Sakaguri\authormark{1}, Masaya Hasegawa\authormark{2}, Yuki Sakurai\authormark{3,4}, Junna Sugiyama\authormark{1}, Nicole Farias\authormark{5},  Charles Hill\authormark{5,6}, Bradley R. Johnson\authormark{7} Kuniaki Konishi\authormark{8}, Akito Kusaka\authormark{1,4,6,9}, Adrian T. Lee\authormark{5,6}, Tomotake Matsumura\authormark{4}, Edward J. Wollack\authormark{10}, and Junji Yumoto\authormark{8}}

\address{\authormark{1} Department of Physics, The University of Tokyo, Bunkyo-ku, Tokyo 113-8654, Japan\\
\authormark{2} High Energy Accelerator Research Organization, Tsukuba, Ibaraki 305-0801, Japan\\
\authormark{3} Graduate school of natural science and technology, Okayama University, Kita-ku, Okayama 700-8530, Japan\\
\authormark{4} Kavli IPMU, The University of Tokyo, Kashiwa, Chiba 277-8583, Japan\\
\authormark{5} Department of Physics, University of California, Berkeley, CA 94720, USA\\
\authormark{6} Physics Divison, Lawrence Berkeley National Laboratory, Berkeley, CA 94720, USA\\
\authormark{7} University of Virginia, Department of Astronomy, Charlottesville, VA 22904, USA\\
\authormark{8} Institute for Photon Science and Technology, The University of Tokyo, Bunkyo-ku, Tokyo 113-0033, Japan\\
\authormark{9} Research Center for the Early Universe, The University of Tokyo, Bunkyo-ku, Tokyo 113-0033, Japan\\
\authormark{10} NASA Goddard Space Flight Center, Greenbelt, MD 20771, USA
}

\email{\authormark{1} kana.sakaguri@phys.s.u-tokyo.ac.jp}



\begin{abstract}
We developed a broadband two-layer anti-reflection (AR) coating for use on a sapphire half-wave plate (HWP) and an alumina infrared (IR) filter for the cosmic microwave background (CMB) polarimetry. Measuring the faint CMB B-mode signals requires maximizing the number of photons reaching the detectors and minimizing spurious polarization due to reflection with an off-axis incident angle. Sapphire and alumina have high refractive indices of 3.1 and are highly reflective without an AR coating.
This paper presents the design, fabrication, quality control, and measured performance of an AR coating using thermally-sprayed mullite and Duroid~5880LZ. This technology enables large optical elements with diameters of \SI{600}{mm}. We also present a newly developed thermography-based nondestructive quality control technique, which is key to assuring good adhesion and preventing delamination when thermal cycling. We demonstrate the average reflectance of about 2.6\% (0.9\%) for two observing bands centered at 90/150 (220/280)\,GHz. 
At room temperature, the average transmittance of a \SI{105}{mm} square test sample at 220/280\,GHz is 83\%, and it will increase to 90\% at \SI{100}{K}, attributed to reduced absorption losses.
Therefore, our developed layering technique has proved effective for 220/280\,GHz applications, particularly in addressing dielectric loss concerns.
This AR coating technology has been deployed in the cryogenic HWP and IR filters of the Simons Array and the Simons observatory experiments and applies to future experiments such as CMB-S4.
\end{abstract}

\section{Introduction}
The cosmic microwave background (CMB) provides various information for understanding the early universe, including inflation~\cite{PhysRevD.60.043504,PhysRevLett.78.2058}. In particular, B-mode polarization, the parity-odd polarization anisotropy pattern, is induced by primordial gravitational waves, gravitational lensing, and foregrounds. The existence of primordial gravitational waves is anticipated, but their measurement has not yet been achieved. The measurement of the B-mode polarization signal is challenging because the signal amplitude is expected to be smaller than \SI{10}{\nano K}~\cite{Bicep21:PhysRevLett.127.151301}.

Intensity-to-Polarization (I-to-P) leakage is one of the major obstacles to CMB polarization measurements. The magnitude of B-mode polarization from primordial gravitational waves is considered seven orders of magnitude fainter compared to the unpolarized CMB temperature anisotropy~\cite{Bicep21:PhysRevLett.127.151301, 2020A&A...641A...6P, Bicep2022:PhysRevD.105.083524}. Minimizing I-to-P leakage caused by the imperfection of the optical elements to a few per mil is crucial. Furthermore, the statistical error has been improved with the growing number of telescope detectors in recent years. Reducing systematic errors to a level lower than the corresponding statistical errors is imperative.

Some of the optical elements, such as a continuously rotating half-wave plate (HWP) and an infrared (IR) filter, have been introduced for high-precision CMB experiments such as the Simons Array (SA) and the Simons Observatory (SO)~\cite{Hill2020hwp,Ali_2020,yamada2023simons}.
An HWP modulates only the polarization signal to mitigate atmospheric 1/f noise and prevents systematics arising from different responses in pairs of polarimeters by rotating continuously~\cite{Takakura_2017,Kusakahwp,Essinger16}. An IR filter blocks infrared radiation that otherwise would heat the cryogenic telescope components. Since IR radiation heats the IR filter, crystalline material, such as alumina, is advantageous for thermal conductivity over a polymer-based filter or a thin membrane. In addition, recent CMB experiments operate at cryogenic temperatures for noise considerations. Implementing these optical elements at cryogenic temperatures is the key to high-precision CMB measurement.

The sapphire and alumina used for HWPs and IR filters have high refractive indices of 3.1~\cite{Inoue:16,Lamb}, which reflects $\sim 40\%$ of light as they are. Regarding the statistical error, there is a concern about a significant decrease in transmittance because each receiver has multiple optical elements.
In addition, high reflectance causes significant I-to-P leakage. The oblique incidence of light into the telescope causes the reflectance to be different for the two orthogonal polarizations, which is a major cause of I-to-P leakage. In particular, polarization from optical elements (e.g., IR filter) on the sky side of the HWP causes a large leakage because the HWP cannot distinguish it from incident polarization.
Highly reflective elements increase systematic and statistical uncertainties. Thus, an anti-reflection (AR) coating is critical for these materials.

There are a few challenging aspects of AR coatings for modern CMB instruments. 
First, applying a coating on 600-mm-diameter alumina surfaces with uniformity of tens of microns is difficult. Second, the SA and SO use dichroic detectors with individual bands with a large, about 30\%, fractional bandwidth~\cite{SA_design, science_goals}. Thus, each AR coating must cover over an octave of bandwidth. Finally, the AR-coated optical elements are cooled down to < \SI{80}{K} to reduce thermal radiation reaching the cryogenic detectors. It creates challenges associated with differential thermal contraction between coating layers.

Some ways to apply AR coatings, like layering dielectrics or machining sub-wavelength structures, have been reported~\cite{Raut2011,Takaku_2021,Golec:22}.
Among these methods, we choose layering dielectrics given the difficulty of machining substrates, which are sapphire or alumina, the required machining time for large-diameter substrates, and the difficulty in applying the high-frequency band.
We establish the fabrication process of the AR coating by introducing new quality control methods. Our AR coating consists of two dielectric layers.
The first layer is thermally sprayed mullite~\cite{Inoue:16}, a ceramic material (Tocalo Corporation)~\cite{TOCALO}. The second layer is Duroid 5880LZ, a composite material (Rogers Corporation)~\cite{ROGERS}, bonded with Epo-Tek 301~\cite{Epoxy}, a two-component room-temperature curing epoxy. We select these materials for their refractive index and coefficient of thermal expansion.
By adjusting the thickness of the coating materials, an AR coating can be optimized for a specific frequency band.

This paper presents the development of broadband multi-layer AR coatings for CMB experiments. Section \ref{sec:design} describes the design. Coating materials are chosen based on the HWP and IR filter requirements. Section \ref{sec:material} presents the materials being fabricated. Section \ref{sec:fabrication and quality control} describes the fabrication process and quality control of the coatings, including tests for delamination. Section \ref{sec:optical characterization} presents the optical characterization of fabricated coatings. Both reflectance and transmittance measurements are described.

\section{Design}
\label{sec:design}
\begin{figure}[tbp]
\begin{minipage}[c]{0.5\linewidth}
\centering
  \includegraphics[keepaspectratio, width=6cm]{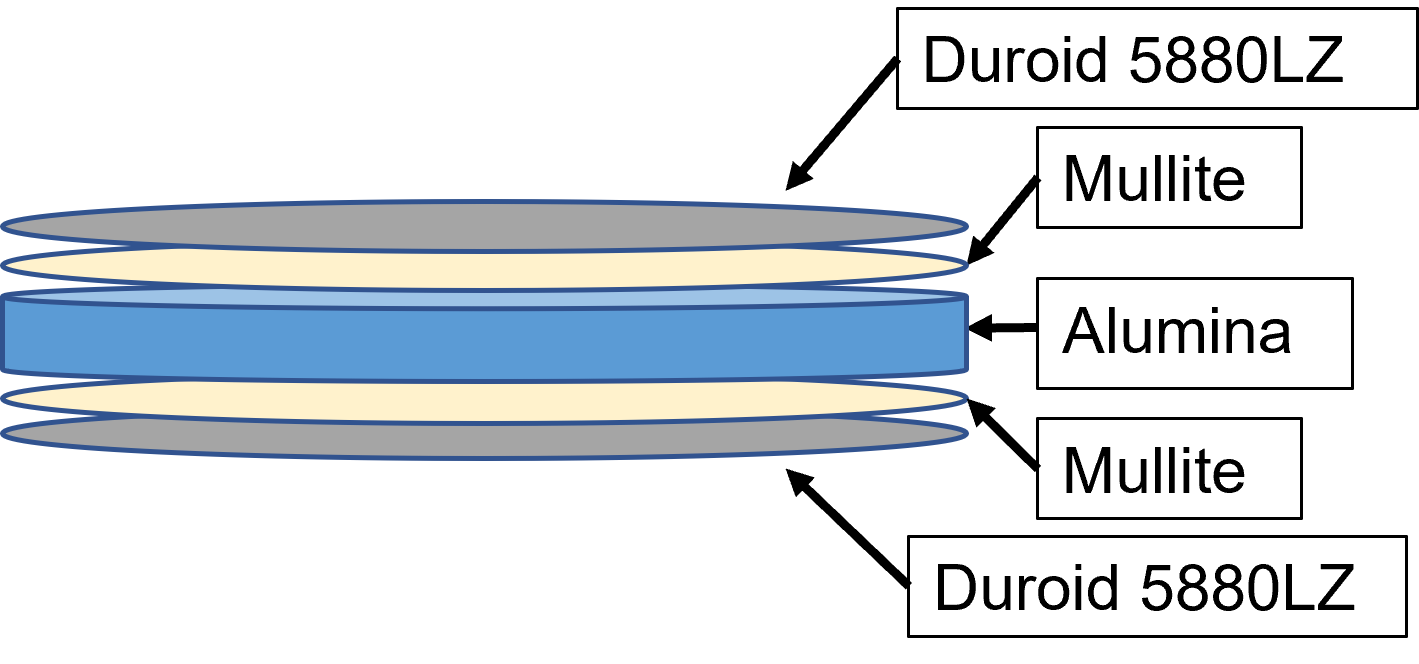}
\end{minipage}
\begin{minipage}[c]{0.5\linewidth}
\centering
  \includegraphics[keepaspectratio, width=6cm]{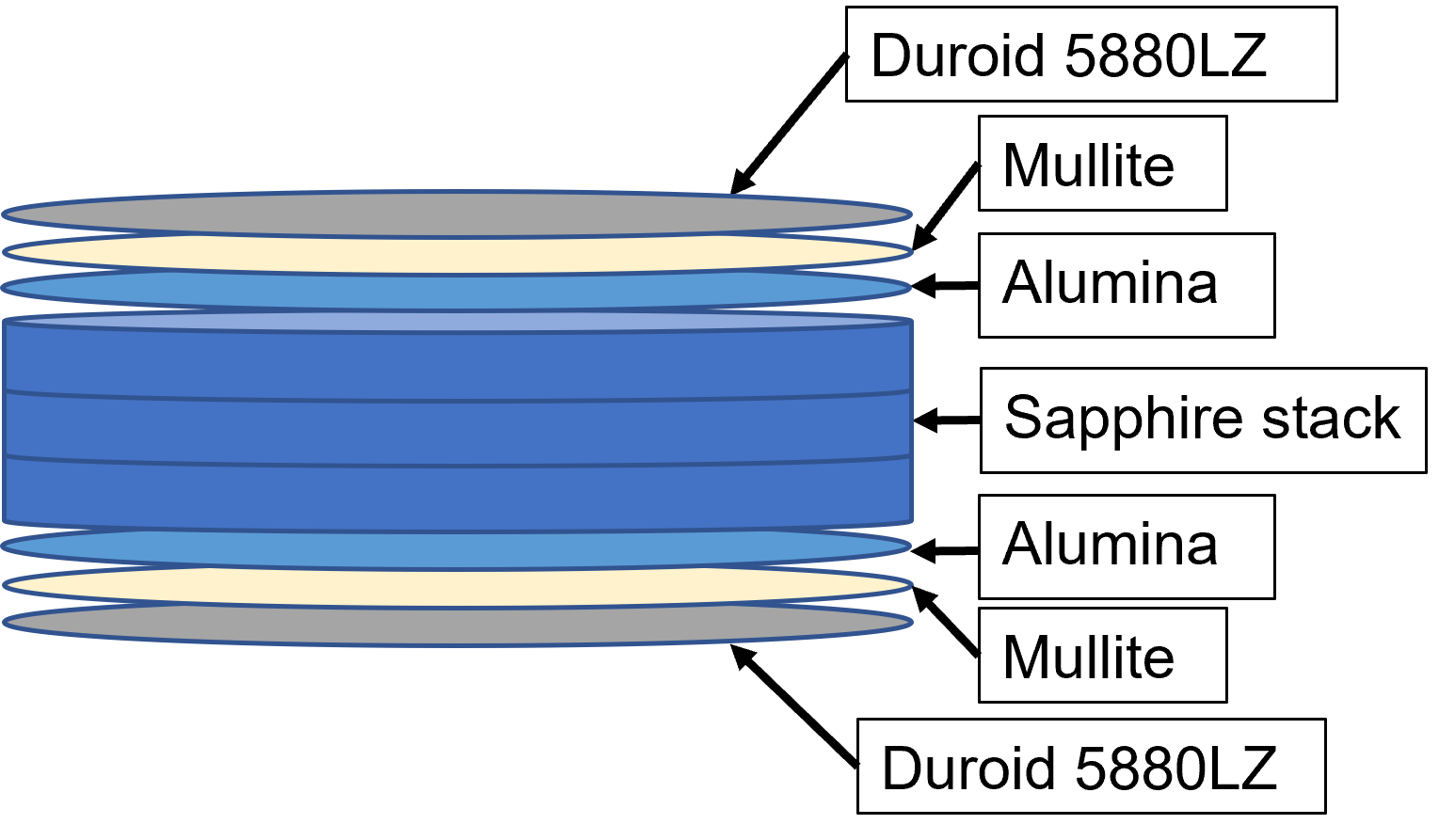}
\end{minipage}
\caption{Left: Schematic view of an alumina IR filter with our AR coating. Reflectance is reduced by layering two dielectric materials with different refractive indexes on both alumina surfaces. The AR coating consists of the thermally-sprayed mullite as the first layer and Duroid 5880LZ as the second layer. Right: Schematic view of a sapphire HWP with the AR coating. The HWP is made of a stack of three sapphire plates. The sapphire stack is sandwiched between one-surface-coated alumina.} 
\label{fig:AR conceptual diagram}
\end{figure}

\begin{figure}[tbp]
    \centering
    \includegraphics[width=8 cm, clip]{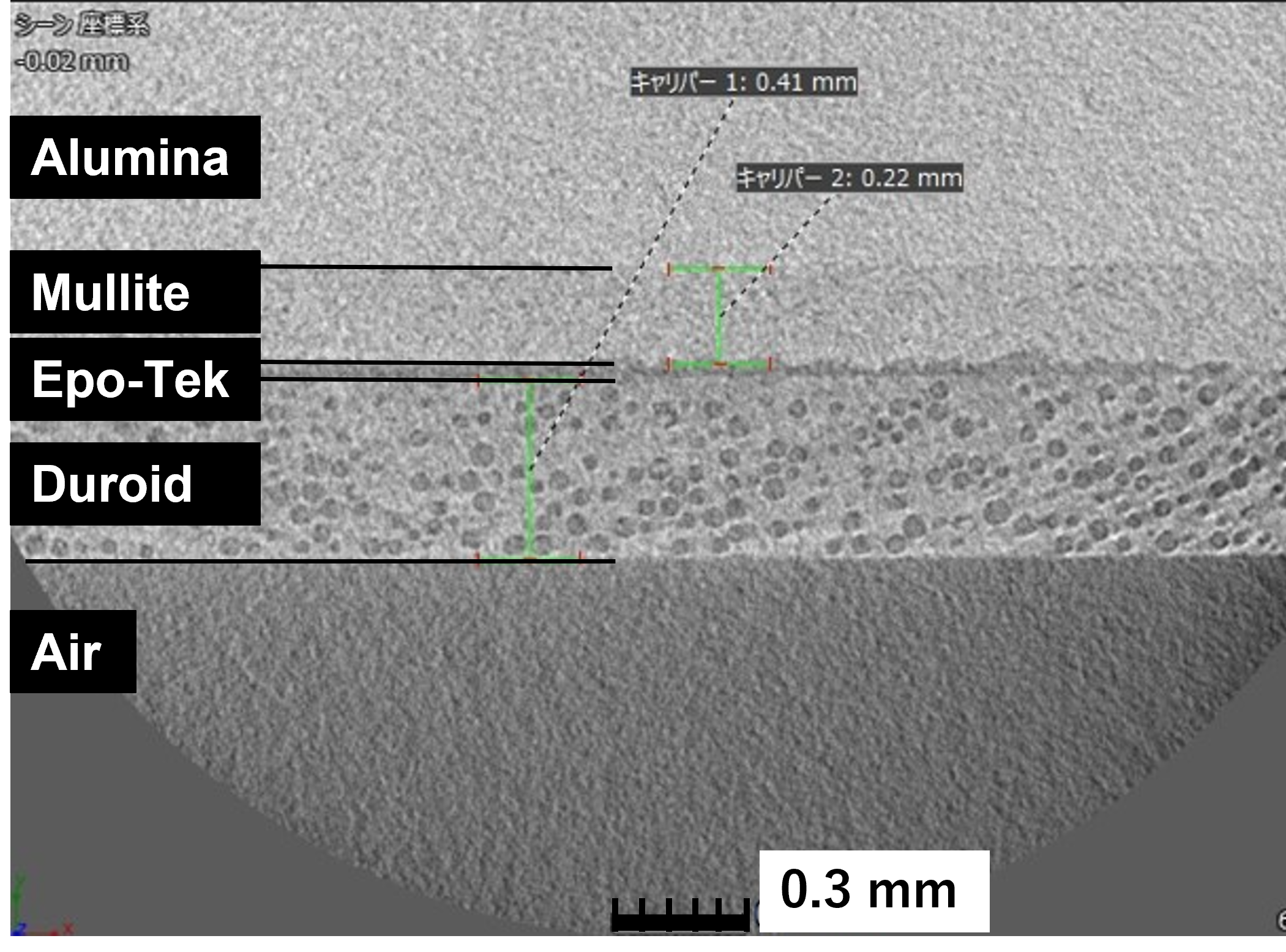}
    \caption{Cross section image of two-layer coating by X-ray CT. The upper layer in the photo is alumina; from there down, there are layers of mullite, Epo-Tek, Duroid, and air. Lines are drawn on the boundary of the layers in the left half.}
    \label{fig:X ray CT}
\end{figure}

We decided on a two-layer coating design for a sapphire HWP and an IR filter. Figure \ref{fig:AR conceptual diagram} shows the schematic of our AR coating. Two layers of mullite and Duroid are coated on alumina surfaces. Details of these materials are written in Section \ref{sec:material}.
Figure \ref{fig:X ray CT} shows the cross-section of a sample taken by X-ray computed tomography (CT). The boundaries of the layers can be seen, and the accuracy of the thickness is \SI{10}{\micro m}. Epo-Tek, a glue between mullite and Duroid, is also uniform and has a thickness of about \SI{40}{\micro m}.


We primarily discuss the AR coating on alumina since the fabrication process is more robust and mature than the coating on sapphire.
As for the sapphire HWP coating, the fabrication process is slightly different. The HWP is made of a stack of three sapphire plates~\cite{Hill:SPIE}. We first fabricate two one-surface-coated aluminas. We then adhere them to both surfaces of the sapphire stack. All the deployable HWP AR coatings we fabricated to date have been achieved this way. However, as described in Sec. \ref{sec:sapphire coating feasibility}, applying the AR coating directly on sapphire is also feasible.

\subsection{HWP and IR filter requirements}
We develop the AR coating for ground-based CMB experiments like the SA. The requirements of the HWP and IR filter are described below:
\begin{itemize}
    \item The HWP diameter is \SI{500}{mm}, and the IR filter diameter is about 500 or 600\,mm~\cite{Hill:SPIE,Ali_2020}. The AR coating needs to be uniform across this large diameter.
    \item The reflectance must be less than a few percent across the detector bands, which typically have $\sim30$\% fractional bandwidth. In addition, the SA and the SO experiments use dichroic detectors. The AR coating should cover these wide bands. We define two frequency bands, Mid-frequency (MF) and Ultra-high-frequency (UHF), based on the experiment bands~\cite{2021,10.1117/12.2312821,abazajian2019cmbs4}.
    We use 78--106\,GHz as MF1, 128--168\,GHz as MF2, 195--256\,GHz as UHF1, and 257--316\,GHz as UHF2 to calculate the average reflectance for each band.
    \item Layers of the  AR coating must not delaminate when thermally cycled to the cryogenic temperature of 4, 40, or \SI{80}{K}~\cite{Ali_2020,Hill2020hwp}. It is thus desirable to have a minimal mismatch of the coefficient of thermal expansion among the AR coating and base materials.
\end{itemize}

\subsection{Material choice}
While several techniques have been developed and reported in the literature~\cite{Golec2020,Nadolski2018,Rosen:13}, repeatable and scalable production of the large-diameter coating remains challenging.
We use a two-layer coating to achieve the required low reflectivity across a broad frequency range.

The coating is optimized by selecting the thicknesses and indices of the two coating layers. As a first-order approximation, the maximum bandwidth of minimal reflection can be achieved when the thickness of each layer is 1/4 wavelength and the indices satisfy: 
\begin{equation}
    n_{\rm{s}} = \frac{n_1^2}{n_2^2},
    \label{eq:optimal index}
\end{equation}
where $n_1$, $n_2$, and $n_s$ are the indices of the first layer, second layer, and sapphire or alumina, respectively~\cite{meyer1989}.
Considering the index, coefficient of thermal expansion, and compressive modulus, we choose coating materials to meet the above-mentioned requirements.

Figure \ref{fig:AR conceptual diagram} schematically shows our AR coatings. Mullite ceramic~\cite{TOCALO} is selected as the first layer, and Duroid 5880LZ~\cite{ROGERS} filled PTFE composite as the second layer.
Mullite had been used as the coating material before~\cite{Inoue:16}, but the combination of mullite and Duroid is first used as a coating for an HWP and a filter for the SA and the SO. Details of these materials are mentioned in Section \ref{sec:material}.

\subsection{Parameter optimization}
Table \ref{tab:AR properties} shows the design thicknesses of the materials used.
The thicknesses are optimized to minimize the average reflectance at normal incidence in each frequency band. Numerical calculations obtain the average reflectance. This prediction is calculated assuming an infinite plane wave incident on a homogeneous dielectric thin layer~\cite{born2000principles}. The 40\,{\textmu}m-thick layer of Epo-Tek~\cite{Epoxy} is placed between two coating layers as glue, and its thickness is included in the average reflectance calculation.
Epo-Tek does not affect the performance because the index of the Epo-Tek is between that of mullite and Duroid, and the thickness is an order of magnitude smaller than that of coating materials.

We fabricate $\SI{105}{mm} \times \SI{105}{mm}$ square samples based on the above calculation. There are differences in coating thickness when we run optical measurements. We hypothesize that the difference between the physical and optical values at the millimeter wave causes the thickness differences. These differences are not a problem in developing an optimal AR coating, as the differences are stable between fabrication runs. The process is repeated by adjusting the thickness of the layers until the optimal coating is obtained.

\begin{table}[tbp]
    \caption{Basic properties of our AR coating materials. The indices, $n$, and thicknesses, $d$, for MF and UHF coatings are shown in this table. Alumina and sapphire are the substrates for the IR filter and the HWP. The alumina thickness varies for optical elements (Table \ref{tab:summary reflectivity after}). Mullite and Duroid are the first and second layers of the coating. The thicknesses are the design values, and the errors represent the production errors. Epo-Tek is a glue between mullite and Duroid. The index is assumed from references~\cite{Epoxy,Munson:17}. The thickness is estimated by X-ray CT measurement (Figure \ref{fig:X ray CT}). All values are cited for room temperature.}
    \centering
    \begin{tabular}{c|c|c|c} \hline
     Material  & $n$    &   $d\,[\si{mm}]$ (MF) & $d\,[\si{mm}]$ (UHF)  \\ \hline
     Alumina   & 3.14   &   --                    & --                      \\ \hline
     Sapphire ordinary axis& 3.05 $\pm$ 0.03\cite{Hill:SPIE} &   \multirow{2}{*}{3.75 $\pm$ 0.01}  & \multirow{2}{*}{1.60 $\pm$ 0.01} \\ \cline{1-2}
     Sapphire extraordinary axis& 3.38 $\pm$ 0.03\cite{Hill:SPIE} &     &  \\ \hline
     Mullite   & 2.52 $\pm$ 0.02\cite{Inoue:16} &   0.212 $\pm$ 0.01     & 0.097 $\pm$ 0.01 \\ \hline
     Duroid    & 1.41 $\pm$ 0.01\cite{ROGERS} &   0.394 $\pm$ 0.01              & 0.183 $\pm$ 0.01 \\ \hline
     Epo-Tek   & 1.7   & 0.04 & 0.04 \\ \hline
    \end{tabular}
    \label{tab:AR properties}
\end{table}

\section{Material Summary}
\label{sec:material}
\subsection{Alumina}
As a substrate, we use A995LD alumina manufactured by NTK ceratec~\cite{NTK}, a low-dielectric-loss type, 99.5\% pure alumina with a refractive index of 3.1 at microwave frequencies. It has desirable material properties for an IR filter: low microwave absorption loss, high IR absorption, and high thermal conductivity at cryogenic temperature~\cite{Inoue:14}.
It is also used for the AR coating substrate of the sapphire HWP (Fig. \ref{fig:AR conceptual diagram}, right). Since both are crystalline ${\rm Al}_2{\rm O}_3$, the indices of sapphire and alumina are well matched as shown in Table \ref{tab:AR properties}, and reflectance due to this minor mismatch is negligible. The thickness of alumina is 2--7\,mm depending on the application. While thinner alumina is desirable, particularly in the UHF band, for lower absorption and scattering, alumina of $<\SI{3}{mm}$ thickness with $\SI{600}{mm}$ diameter does require handling with care.

\subsection{Sapphire}
We use $\alpha$-cut sapphire plates manufactured at Guizhou Hoatian Optoelectronics Technology (GHTOT)~\cite{GHTOT} as a cryogenic birefringent substrate for HWPs~\cite{Hill2020hwp}. We stack three sapphire plates to achieve broadband polarization modulation efficiency~\cite{Pancharatnam1955}. As shown in Table \ref{tab:AR properties}, sapphire has a large differential index and small loss tangent~\cite{Hill:SPIE}. The refractive index of the ordinary (extraordinary) axis is 3.1 (3.4). The MF (UHF) HWP consists of three 3.75-mm-thick (1.6-mm-thick) sapphire plates for optimum modulation efficiency across the frequency band and practicality~\cite{Hill:SPIE, Pancharatnam1955}.

\subsection{Mullite}
We use mullite as the first layer coating material because the coefficient of thermal expansion, which is $\sim 10^{-6}$/K at room temperature, is close to that of alumina.
Mullite ($3{\rm Al}_2{\rm O}_32{\rm SiO}_2$) is a silicate with a refractive index of 2.52. It is a porous ceramic material that can be applied to alumina using the thermal-spraying method offered by Tocalo Corporation~\cite{TOCALO}. 
This process uses an established and available technology and can be fabricated to a high precision of \SI{10}{\micro m} in thickness. Mullite does not cause delamination under cryogenic conditions because it has a coefficient of thermal expansion matched that of alumina~\cite{Inoue:16}.  
Mullite has a surface roughness due to its asperity structure. The design value of the thickness is 
\SI{0.212}{mm} at MF and \SI{0.097}{mm} at UHF based on X-ray CT and optical measurements.

\subsection{Duroid}
We use Duroid 5880LZ manufactured by Rogers Corporation as the second layer. 
Duroid 5880LZ is a matrix of polytetrafluoroethylene (PTFE) loaded with a filler of \SI{50}{\micro m} diameter hollow, nitrogen-filled, aluminosilicate microspheres with a refractive index of 1.41 at millimeter wave~\cite{ROGERS}.
The manufactured thickness of the Duroid sheet is \SI{500}{\micro m}, and we machine it down to 385$\pm$10\,{\textmu}m for MF and 155$\pm$10\,{\textmu}m for UHF at Suzuno Giken~\cite{Suzuno}.

\subsection{Epo-Tek}
We use Epo-Tek 301-2, an aerosolized optical epoxy manufactured at Epoxy Technology~\cite{Epoxy} as a glue to apply the Duroid sheet to the mullite layer.
We assume the refractive index of Epo-Tek is 1.7 at the millimeter wavelengths, which is between 1.53 at \SI{600}{\nano m} and 1.9 of the static value~\cite{Munson:17,Epoxy}. The difference in reflectance due to these differences is less than 0.4\%, which indicates that Epo-Tek does not compromise the optical performance.
The thickness of the Epo-Tek layer is 40\,{\textmu}m (Figure \ref{fig:X ray CT}).

\section{Fabrication and quality control}
\label{sec:fabrication and quality control}

\subsection{Fabrication}
\begin{figure}[tbp]
    \centering
    \includegraphics[width=12 cm, clip]{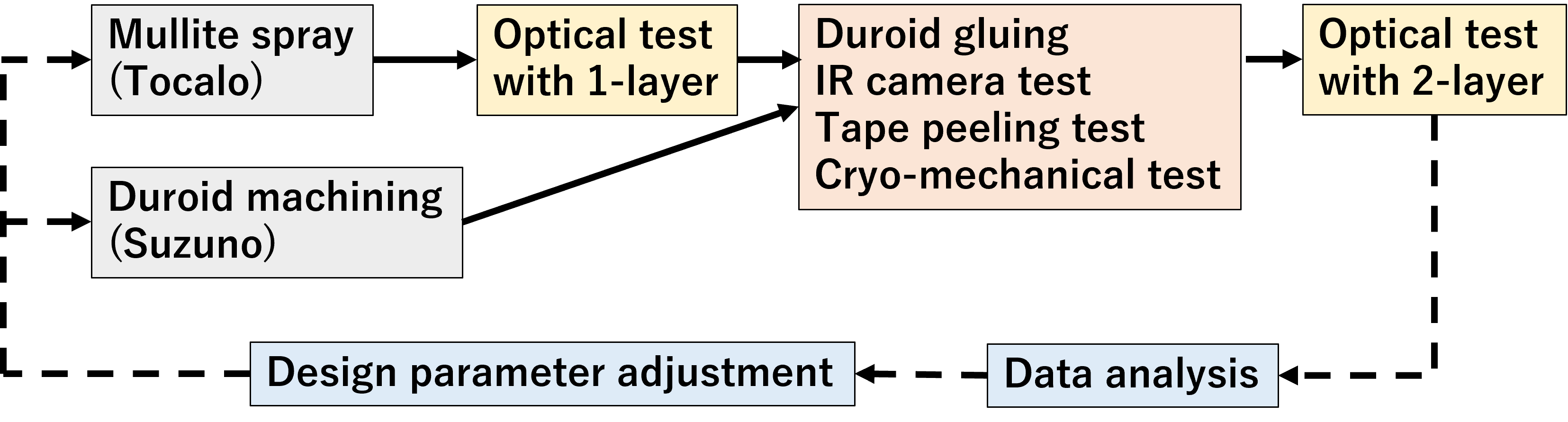}
    \caption{Schematic of the fabrication and R\&D process. The solid arrows show the fabrication process of one coating. The dotted arrows indicate that a new sample is being made. This process is repeated by adjusting the thickness of the layers until the optimal reflectivity is obtained.}
    \label{fig:Fab schematic}
\end{figure}

Figure \ref{fig:Fab schematic} shows the fabrication process. As a small test sample, a \SI{105}{mm} square of alumina with a thickness of \SI{4}{mm} is used. In the process, we check the optical performance every time we add a layer to confirm the properties of each layer.

The first layer, mullite, is thermally sprayed on both alumina surfaces~\cite{Inoue:16}.
Atmospheric plasma spraying by Tocalo Corporation enables deposition with an accuracy of 10\,{\textmu}m~\cite{TOCALO}.
The second layer of Duroid is glued on with 40\,{\textmu}m-thick Epo-Tek.
To compute the accurate model during simulation, the reflectance is calculated with three layers, including the thin layer of Epo-Tek.
Epo-Tek is applied to the Duroid side and uniformly bonded to the mullite. We first use a roller to remove the air bubbles from the bond interfaces, and then the assembly is vacuum bagged to apply a uniform load while curing the epoxy.
After gluing, we dice the Duroid sheet to prevent peeling due to heat shrinkage using a box cutter into $40 \times 40$\,mm square islands for stress relief, while the mullite layer is not diced after applying the Duroid sheet. The cut widths are typically less than 100\,{\textmu}m, and no optical degradation is expected.

The optical performance of the test samples is measured to determine the effective thicknesses of each layer. The thickness of each layer differs from the design value in optical measurements. Duroid appears to be $\sim$ 20\,{\textmu}m thicker, and mullite seems to be $\sim$ 40\,{\textmu}m thinner than their design values. Adjusting and refabricating a sample (Figure \ref{fig:Fab schematic}) gives the intended optical performance. Section \ref{sec:optical characterization} describes the optical setup and performance.

We fabricate a large-diameter coating in the same way as the test sample after determining the optimal thickness from the test samples. We fabricate five sets of coatings for MF and three coatings for UHF.
Table \ref{tab:summary reflectivity after} shows the properties of these coatings.
The thicknesses of alumina substrates depend on the optical element, and they do not affect the performance of the coatings. The optical performance of each coating will be discussed in Section \ref{sec:optical characterization}.

\begin{table}[bp]
\centering
\caption{Summary of the coatings. Numbers one and two are used for the \textsc{Polarbear}-2b telescope~\cite{SA_design,howe2018design}.
Coatings numbered three to five, seven, and eight are developed for the SO small aperture telescope~\cite{Ali_2020}.
The sixth coating is fabricated for the SO large aperture telescope~\cite{Zhu_2021}. S1 and S2 are 105\,mm $\times$ 105\,mm test samples fabricated with final thicknesses shown in Table \ref{tab:AR properties}. $d_s$ shows the thickness of the alumina substrate. The reflectivity given for the large-diameter coatings is averaged over all the measurement points. The average reflectance is calculated for the entire band on each dichroic detector response's low and high frequency sides, respectively.}
\label{tab:summary reflectivity after}
\begin{tabular}{c|c|c|c|c|c|c|c} \hline
\multirow{2}{*}{No.}   & \multirow{2}{*}{diameter [mm]} & \multirow{2}{*}{element}     & \multirow{2}{*}{$d_s$ [mm]} & \multirow{2}{*}{band} &\multicolumn{3}{c}{reflectivity [\%]} \\ \cline{6-8}
                       &                                &                              &                             &                        & low & high & Ave. \\ \hline\hline
1      & 530           & \SI{40}{K} filter & 3          & MF   & 1.1      & 2.0       & 1.7       \\
2      & 505           & HWP         & 4          & MF   & 1.3      & 4.3       & 3.0       \\
3      & 600           & \SI{40}{K} filter & 3          & MF   & 1.7      & 1.6       & 1.6       \\
4      & 600           & \SI{4}{K} filter  & 3          & MF   & 0.8      & 3.8       & 2.6       \\
5      & 505           & HWP         & 4          & MF   & 1.2      & 6.1       & 4.0       \\
6      & 396           & \SI{80}{K} filter & 7          & UHF  & 0.22     & 0.65      & 0.43      \\
7      & 600           & \SI{40}{K} filter & 3          & UHF  & 1.51     & 1.65      & 1.58      \\
8      & 600           & \SI{4}{K} filter  & 3          & UHF  & 1.39     & 0.75      & 1.08      \\ \hline\hline
S1  & 105\,mm $\times$ 105\,mm &             & 4          & MF   & 1.7      & 3.4       & 2.7       \\
S2  & 105\,mm $\times$ 105\,mm &             & 4          & UHF  & 0.38     & 0.87      & 0.62      \\ \hline
\end{tabular}
\end{table}

\subsection{Quality control}
The key to control quality is uniform and robust coatings. The Duroid sheets must be securely coated to the mullite surface to avoid air gaps, which would cause delamination in vacuum and cryogenic conditions. The fabricated coatings must be tested non-destructively. We developed three tests: an IR camera test, a tape peeling test, and a cryo-mechanical test. The IR camera test is crucial in establishing the fabrication process. Our testing found that all coatings that passed the IR camera test also successfully cleared the cryo-mechanical tests. In addition, optical characterization is essential to evaluate the optical performance of AR coatings. We will discuss the optical tests in Section \ref{sec:optical characterization}.

\begin{figure}[tbp]
\centering
  \includegraphics[width=12 cm,
  clip]{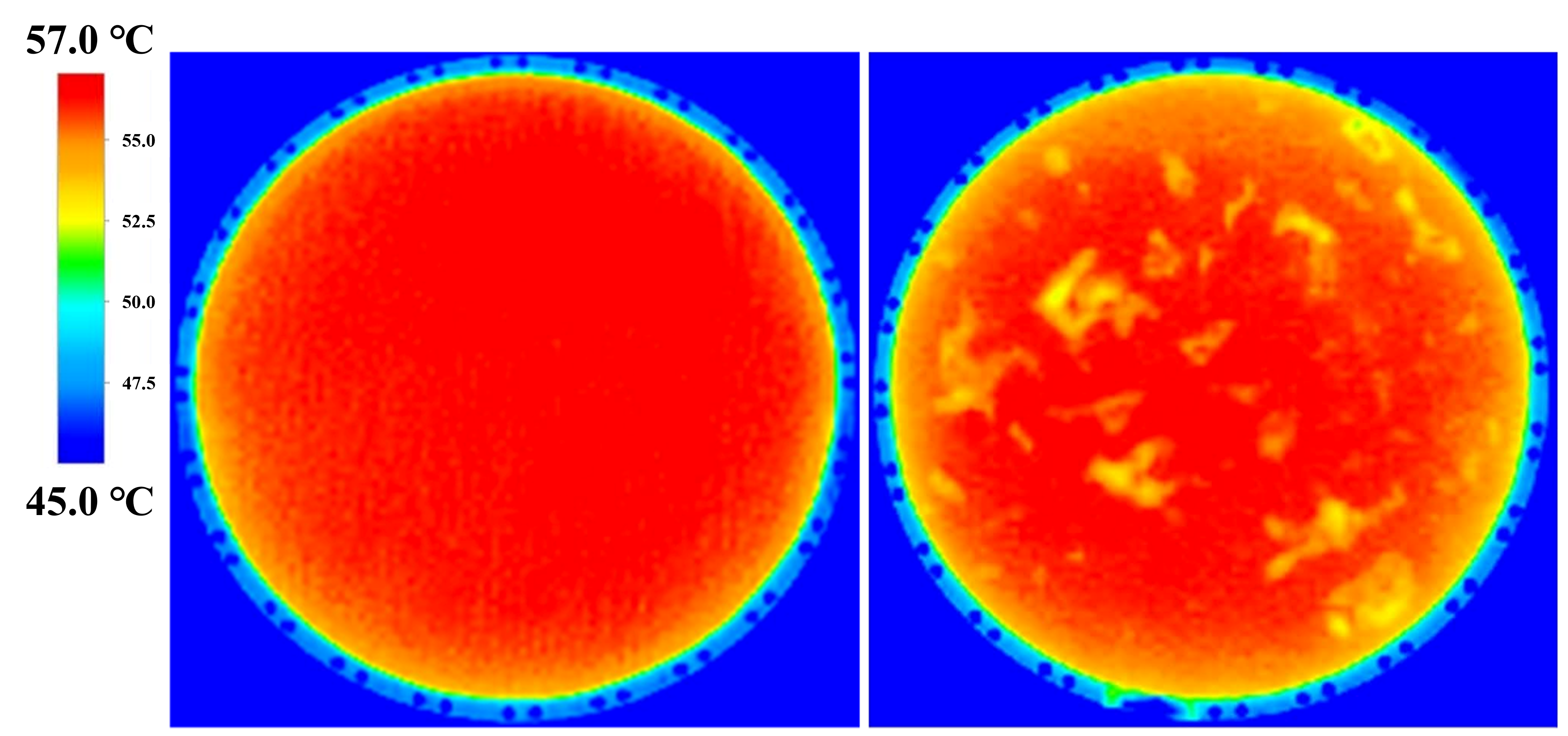}
\caption{ Image of the Duroid surface using an Infrared camera for a good (left) and a failed (right) sample. The diameter is \SI{505}{mm}. The delamination areas are observed as cold spots due to poor thermal conductivity.}
\label{fig:IRcamera}
\end{figure}

We use an IR camera to look for areas of low thermal conductivity (i.e., areas with air gaps) to diagnose whether the adhesion is successful.
The procedure is to warm the fabricated sample uniformly to \SI{60}{\degreeCelsius} in a thermostatic chamber~\cite{Maxwell2007}. The surface is then photographed using an IR camera. 
Figure \ref{fig:IRcamera} shows the example of the IR view for both good and delaminated samples. 
If the sample is not well adhered, the surface will show cold spots, as shown in the right panel of Figure \ref{fig:IRcamera}.

We employ a tape peeling test after the IR camera test.
The surface of the coating is pulled using tape and checked to see if the adhesive strength is sufficient. We select a tape with appropriate adhesion properties because using too much adhesion could cause the surface to be torn. We use curing tape~\cite{DIATEX} as it does not damage the surface and has good consistency of results, as shown by the IR camera test.

We perform a cooldown test as a cryo-mechanical test to ensure the fabricated samples do not delaminate when cooled. We cool the large-diameter coatings more than two times to \SI{30}{K}.
No delamination after the cooldown occurred for samples that had passed the IR and tape test.


\subsection{Sapphire coating feasibility}
\label{sec:sapphire coating feasibility}
\begin{figure}[tbp]
    \centering
    \includegraphics[width=6 cm, clip]{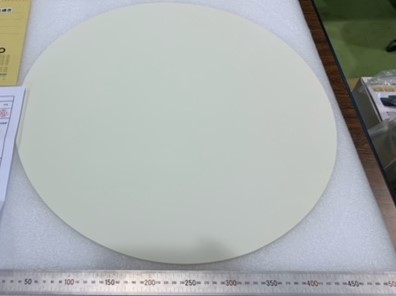}
    \caption{Sapphire directly coated with mullite. The diameter is \SI{505}{mm}.}
    \label{fig:Sapphire caoting}
\end{figure}

It is challenging to coat directly on a sapphire surface. We can not apply the same approach to directly coat sapphire with mullite because sapphire is vulnerable to cracking during the required pre-heating step due to its internal stress.
We can prevent this damage by annealing the sapphire, which relaxes the internal stress \footnote{The annealing process is performed by KYOCERA Cooperation~\cite{Kyocera}.}. The mullite coating process on annealed sapphire follows the same process applied to alumina. The mullite layer has been successfully applied directly to sapphire with diameters of 100, 390, and \SI{505}{mm} (Figure \ref{fig:Sapphire caoting}). R\&D continues on large-diameter coatings.

\section{Optical characterization}
\label{sec:optical characterization}
\subsection{Reflectance}
\begin{figure}[tbp]
\begin{minipage}[c]{0.5\linewidth}
\centering
  \includegraphics[keepaspectratio, width=6cm]{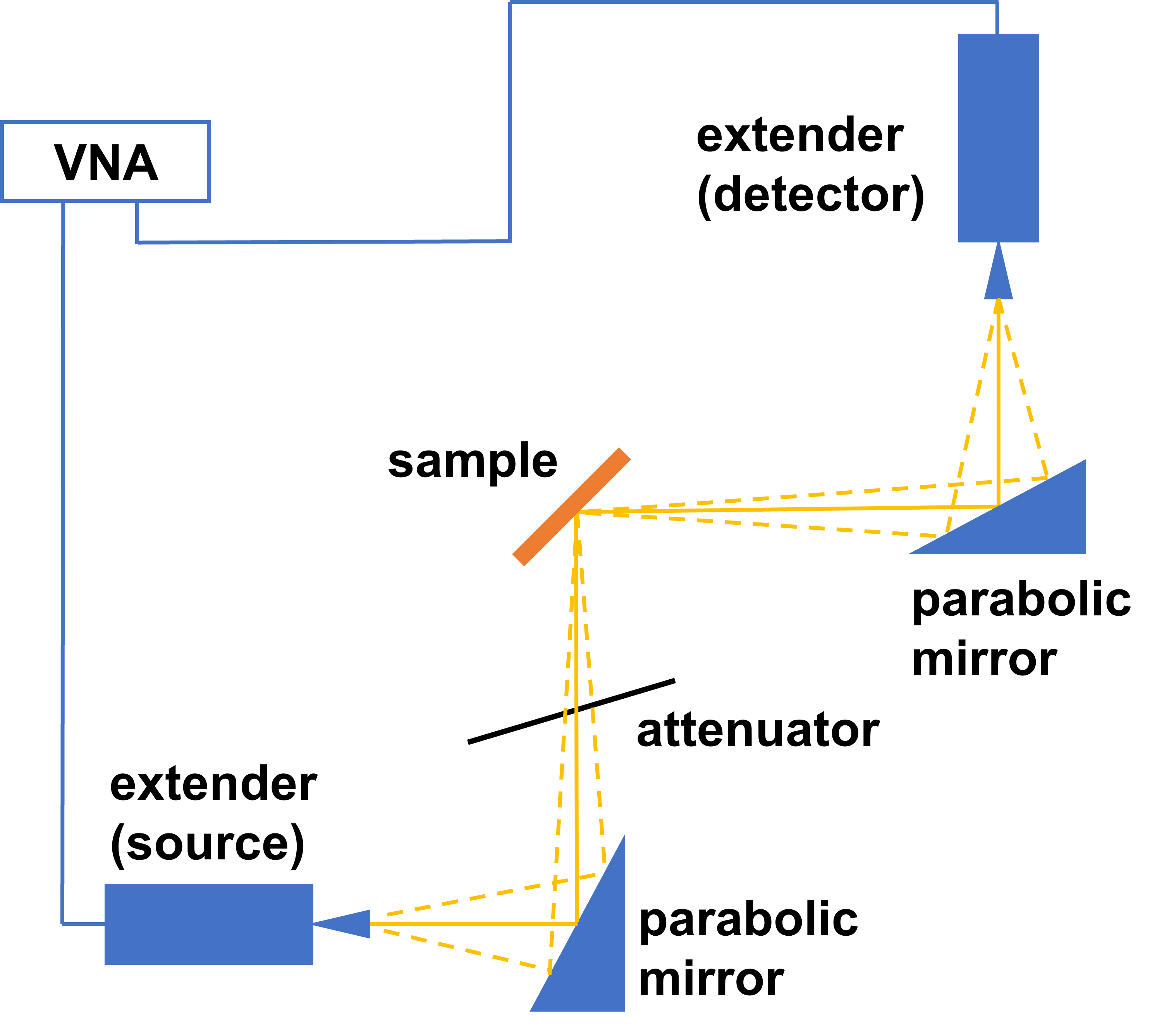}
\end{minipage}
\begin{minipage}[c]{0.5\linewidth}
\centering
  \includegraphics[keepaspectratio, width=6cm]{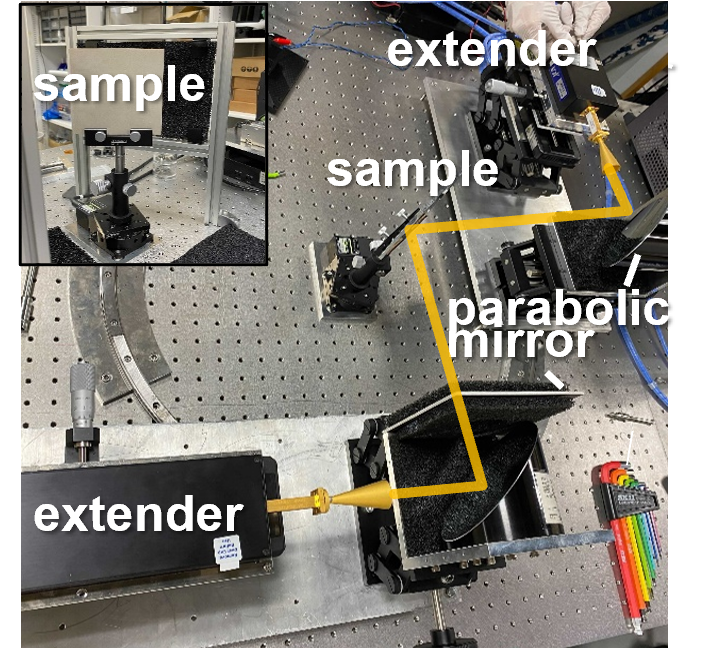}
\end{minipage}
\caption{Left: Schematic of the reflectance measurement setup. The yellow lines indicate the optical path. A parabolic mirror reflects the electromagnetic wave output from the extender and enters the sample as S-polarized light. The parabolic mirror reflects the light reflected by the sample on the other side and enters the extender on the receiving side. Right: Picture of the measurement setup for reflection. The top left photo shows the vertical view of the sample.} 
\label{fig:setup reflection}
\end{figure}

We perform optical tests on the samples that pass the quality control. The optical performance of the coatings is evaluated with a Keysight N5222B vector network analyzer (VNA)~\cite{keysight}. We measure the reflectivity at an incident angle of 45 degrees and the transmissivity at 55--300\,GHz according to each observing band.
Figure \ref{fig:setup reflection} shows the schematic of the reflectance measurement setup. A parabolic mirror reflects the output from the source side extender and focuses on the sample as S-polarized light. The sample reflects it and enters the detector side extender. This allows the reflectance at 45-degree incidence to be measured. The systematic errors in the measurement are estimated using an alumina slab as a reference, for which the refractive index and thickness can be measured well. We first measure the slab and calculate the residuals.

Using the obtained reflectivity, or $|S_{21}|^2$, which is the coupling between the amplitude of the source port and the detection port, we fit the index and the thickness of the coating materials at a 45-degree incidence (Figure \ref{fig:setup reflection}).
Once the thicknesses and indices of the layers are determined, we use them to calculate the reflectance and transmittance at any angle of incidence.
The on-axis reflectivity is calculated by substituting the fit values into a model with a zero-degree incident angle.

\begin{figure}[tbp]
\begin{minipage}[b]{0.5\linewidth}
\centering
  \includegraphics[keepaspectratio, scale=0.4]{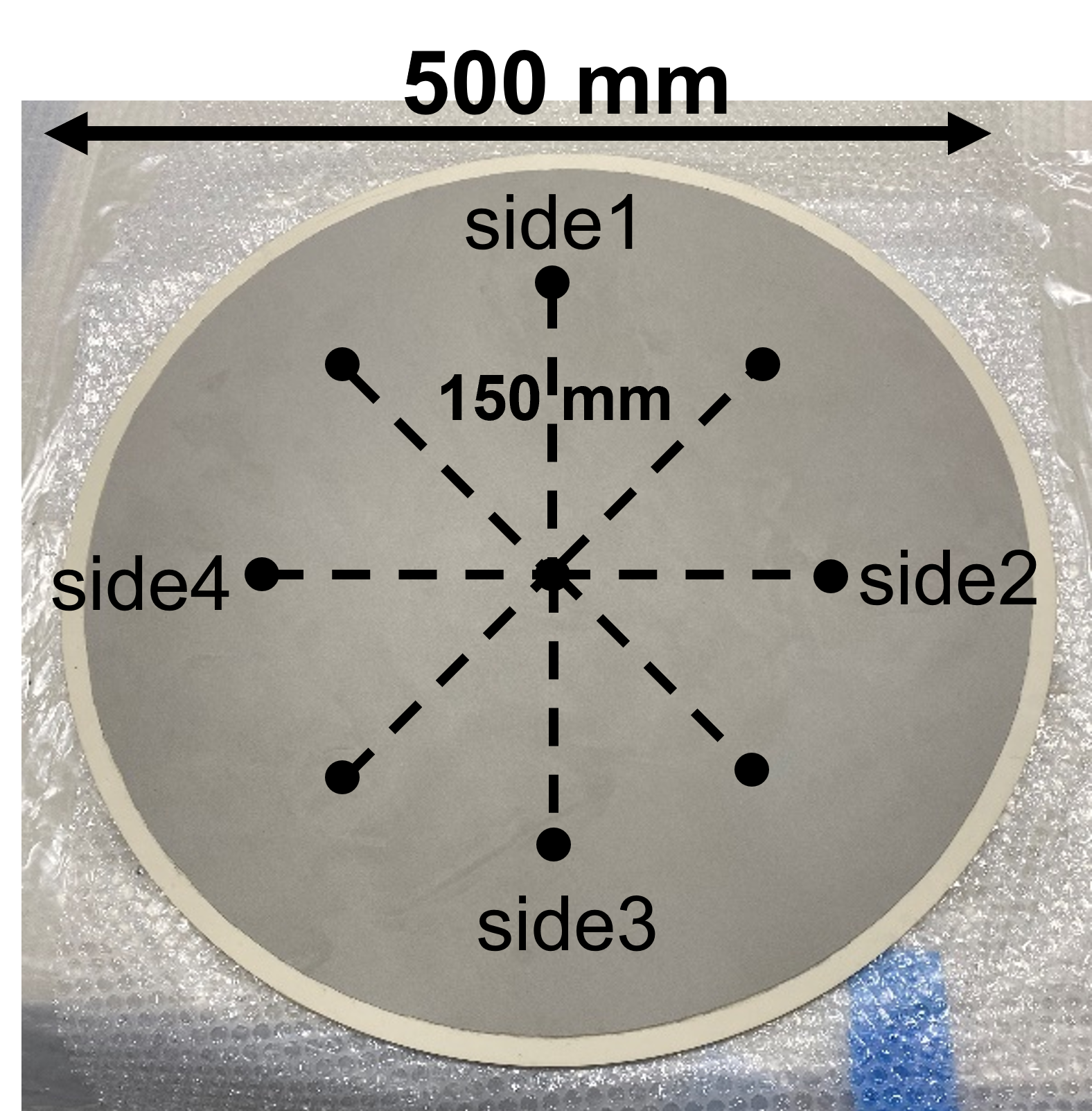}
\end{minipage}
\begin{minipage}[b]{0.5\linewidth}
\centering
  \includegraphics[keepaspectratio, scale=0.4]{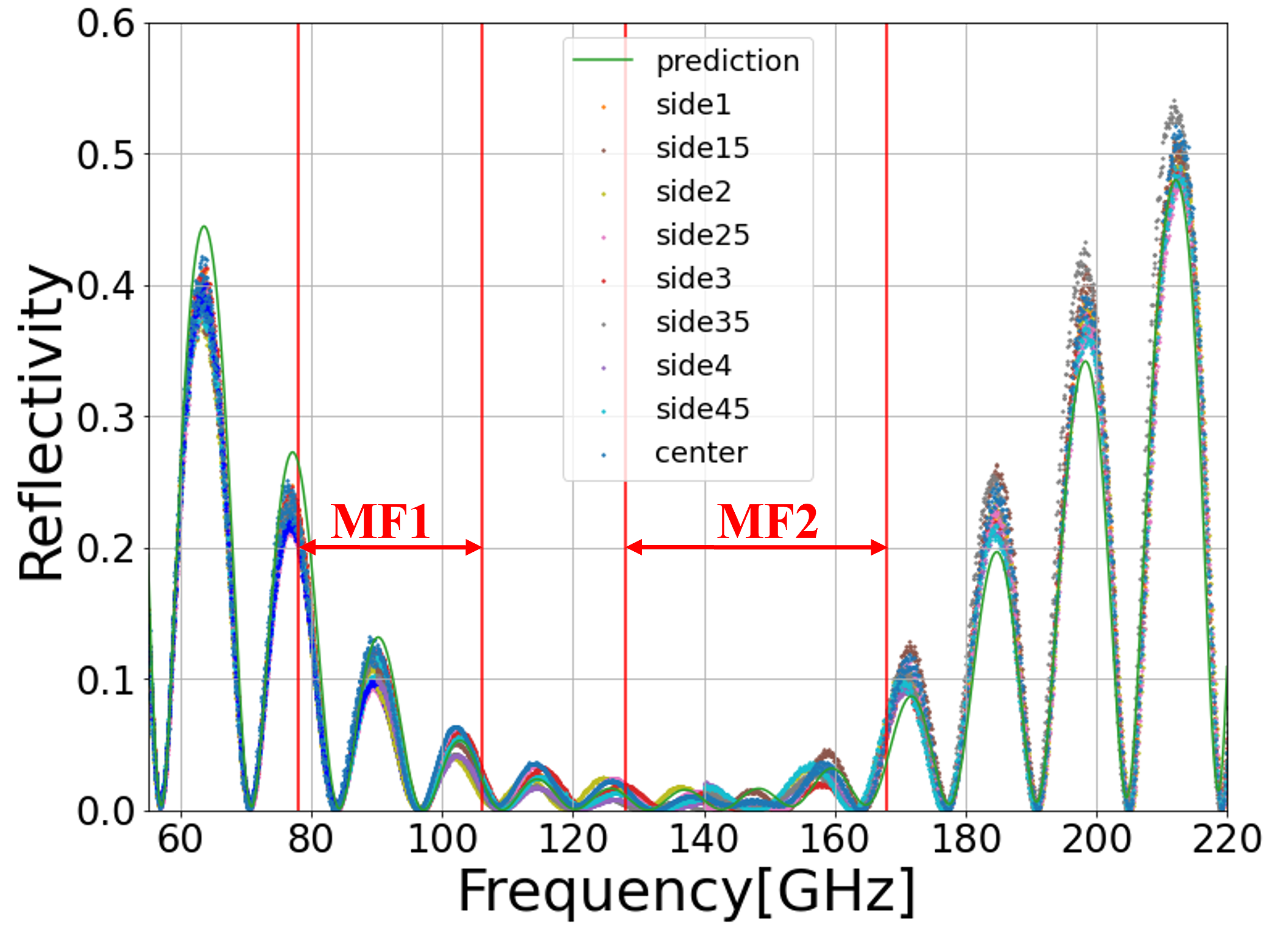}
\end{minipage}
\caption{Left: IR filter with AR coating. Black points show the measurement points. The center and 8 points around the perimeter were measured, each \SI{150}{mm} away from the center. Right: Measured reflectivity of item No. 3 (raw data at an incidence angle of 45 degrees). The red lines show the frequency bands we assume to calculate the average reflectance. Each data point corresponds to the point in the left figure. Side 15 (25/35/45) means the point between side 1 (2/3/4) and side 2 (3/4/1).} 
\label{fig:large samples}
\end{figure}

\begin{figure}[tbp]
\begin{minipage}[c]{0.5\linewidth}
\centering
  \includegraphics[keepaspectratio, width=7cm]{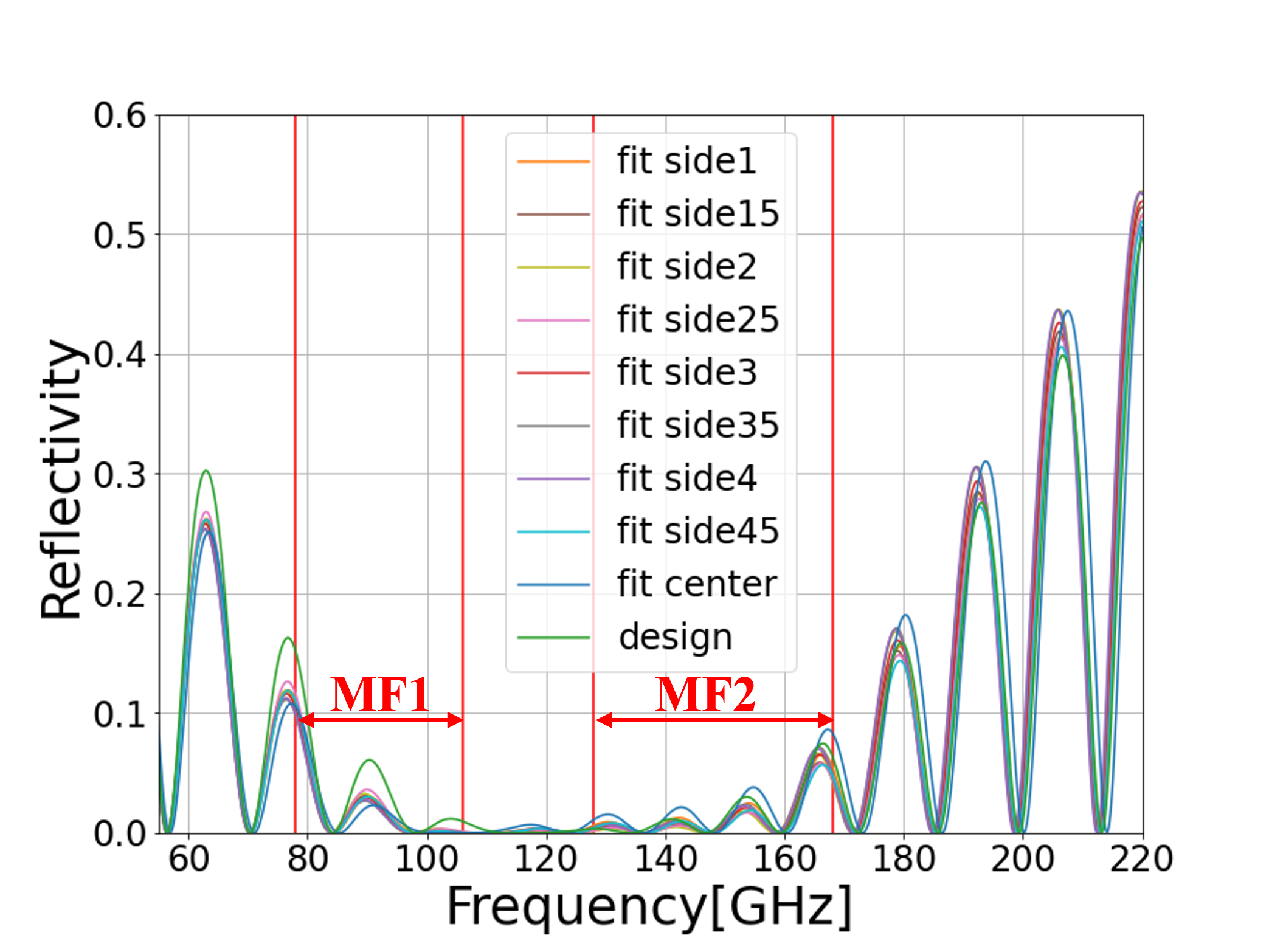}
\end{minipage}
\begin{minipage}[c]{0.5\linewidth}
\centering
  \includegraphics[keepaspectratio, width=7cm]{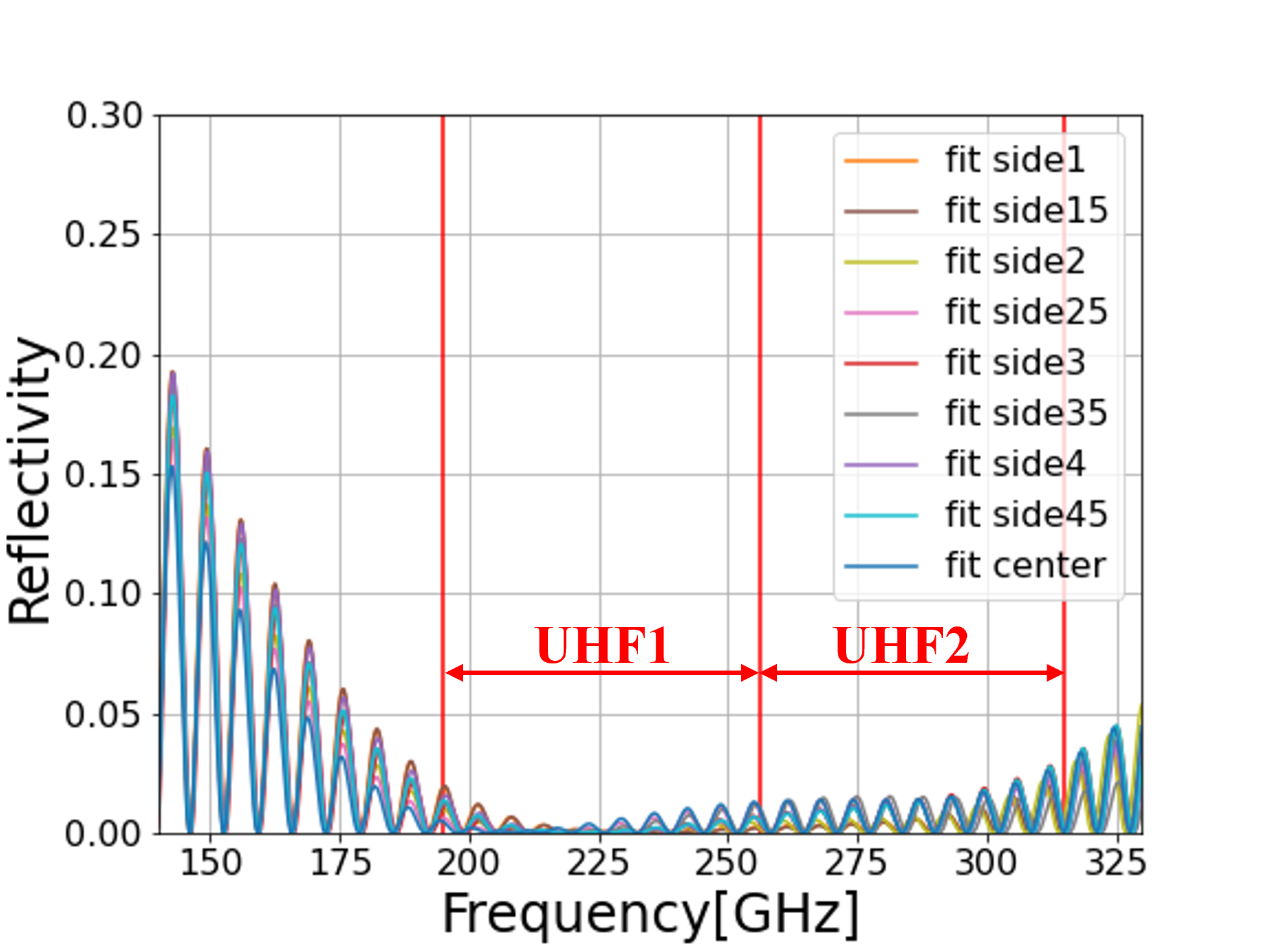}
\end{minipage}
\caption{Left: On-axis projected performance of an MF coating (No.~3). Right: On-axis projected performance of a UHF coating (No. 6).} 
\label{fig:Lager coating On-axis}
\end{figure}

Table \ref{tab:summary reflectivity after} shows the summary of the reflectivity of all fabricated coatings.
The band average reflectivity achieved is 1.7\%/3.4\% in the MF sample and 0.38\%/0.87\% in the UHF sample~\cite{Sakaguri2022}.
The larger coatings, with a diameter of \SI{600}{mm}, fabricated for practical use exhibit performance comparable to the test samples. In addition, the reflectance is uniform across the large-diameter coatings. Figure \ref{fig:large samples} shows one of the large coatings and the raw data of this measurement. Nine locations are measured to check the uniformity. Since the figure shows the off-axis results, the final average reflectance is calculated by extrapolating the fit results of all the measurement points to the on-axis incidence.
Figure \ref{fig:Lager coating On-axis} shows the on-axis performance of coatings on two large surfaces. The reflectance of the coatings on large surfaces is similar to the reflectance of the test samples within errors of a fabrication tolerance. In addition, the leakage, which is the difference in reflectance between the S and P polarization, is calculated for the coatings for the SA and the SO receivers. Values of the leakage due to oblique incidence (at an incident angle of about 17 degrees, which is reasonable in SO-like telescopes~\cite{Ali_2020}) obtained from these fits satisfy the requirement that the leakage is less than 0.2\% for the \SI{40}{K} filter, for example.

\begin{figure}[tbp]
    \centering
    \includegraphics[width=8 cm, clip]{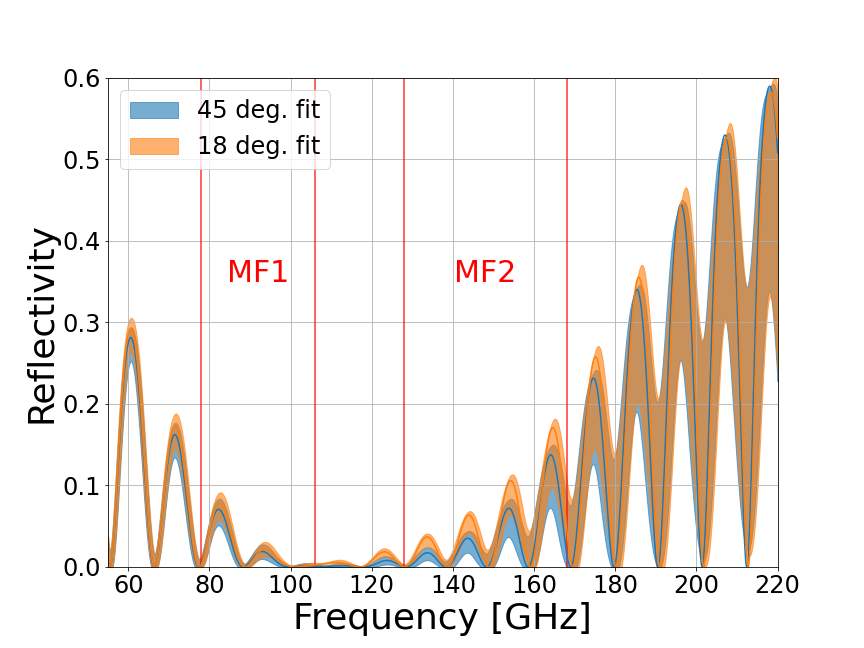}
    \caption{Comparison of calculated on-axis projected reflectance based on measurements at different angles of incidence. The fitted results of measurements at 18 (orange) and 45 (blue) degrees represent the results of extrapolation to vertical incidence, respectively. Each region shows the 1-$\sigma$ region. Two measurements are consistent with the errors. Note that this is a coating during the R\&D process, and its performance is not finalized.}
    \label{fig:18 45deg comparison}
\end{figure}

We confirmed validity through consistency between results from two different angles of incidence. One is to measure the reflectivity of a sample at an angle of incidence of 18 degrees. Reflectance is measured in the same way as at 45 degrees, and then the thickness and index of each layer are fitted. The difference in thickness between the 45-degree and 18-degree measurements is within \SI{10}{\micro m}. As for the index, the difference in the fit results is <2\%. Figure \ref{fig:18 45deg comparison} shows the on-axis performance extrapolating these fitted values to zero-degree incidence. This coating is fabricated during the R\&D process. The blue/orange region shows the 1-$\sigma$ region of the measured values at an incident of 45/18 degrees. These two measurements are consistent.
The other way is to measure the phase of $S_{21}$ parameter. Both the absolute and phase values of $S_{21}$ are measured to resolve the degeneracy of fitted thicknesses in some samples. The measured phase is consistent with the model.

\subsection{Transmittance}
We measure the transmission for some samples to estimate the AR coating's loss tangent, $\tan\delta$. One can reasonably expect that the loss properties of the material are rather stable between samples; we thus consider that transmission measurements of only a limited number of samples suffice for our needs. Absorption increases as the frequency increases, and it is necessary to estimate them, especially for the UHF samples with high absorption.

\begin{figure}[tbp]
\begin{minipage}[c]{0.5\linewidth}
\centering
  \includegraphics[keepaspectratio, width=6cm]{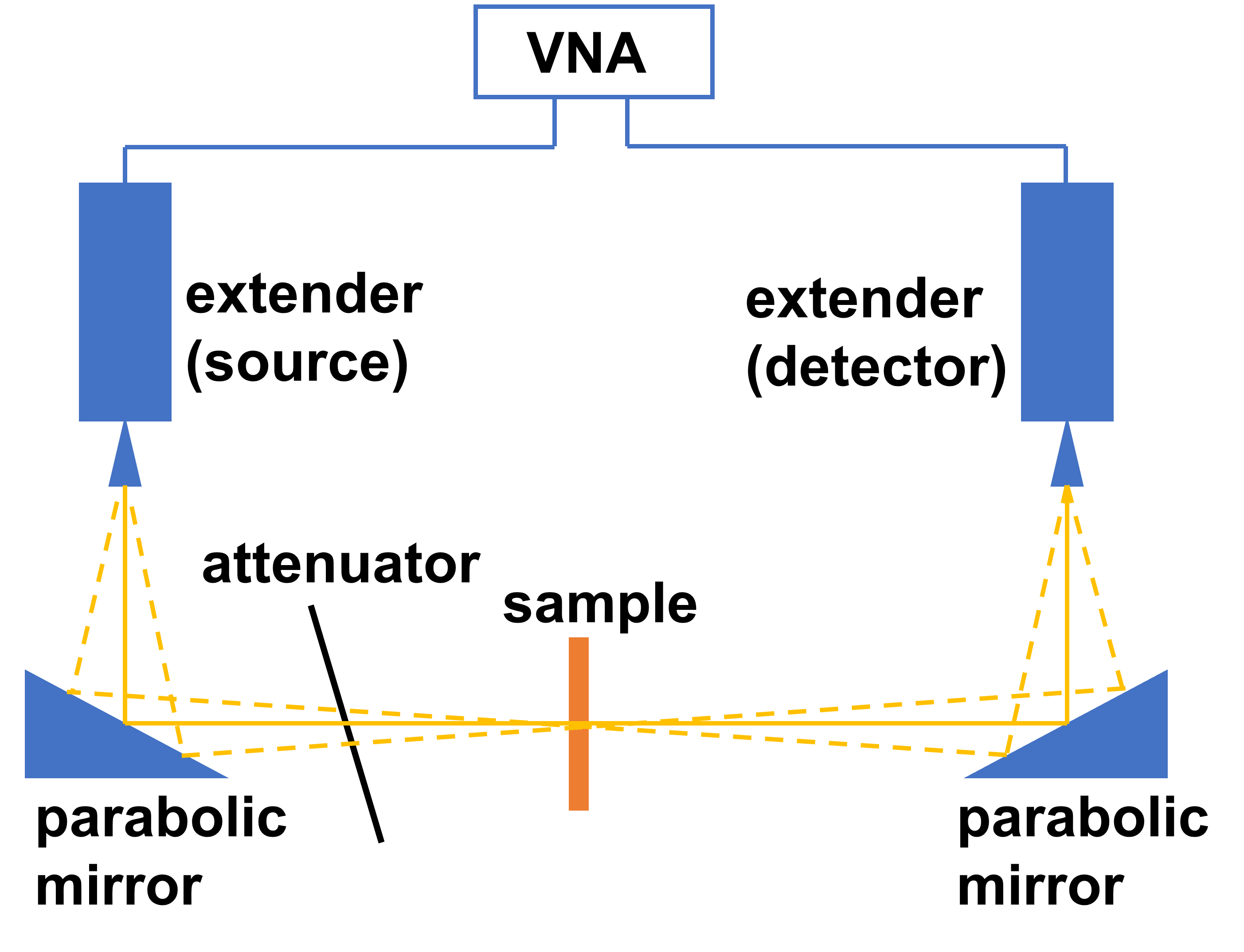}
\end{minipage}
\begin{minipage}[c]{0.5\linewidth}
\centering
  \includegraphics[keepaspectratio, width=6cm]{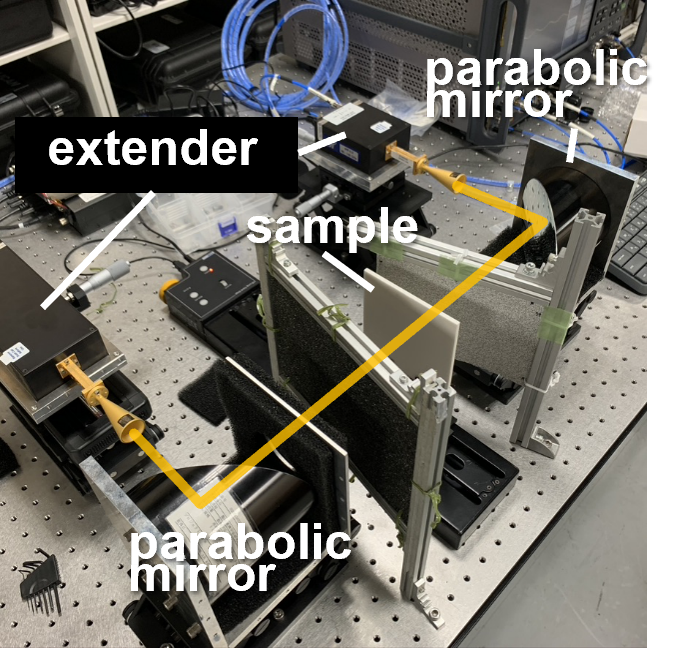}
\end{minipage}
\caption{Left: Schematic of the transmittance measurement setup. The yellow lines indicate the optical path. The setup is almost the same as the reflection measurement. Right: Picture of the measurement setup for transmission. The sample is tilted a few degrees to avoid standing waves.} 
\label{fig:setup transmission}
\end{figure}

Figure \ref{fig:setup transmission} shows the schematics and setup of transmittance measurements for test samples. The measurement method is the same as that for reflectance measurements. The angle of incidence is zero degrees, but the sample is tilted about five degrees to reduce standing waves. The average transmittance of a 105\,mm $\times$ 105\,mm UHF sample (S2 in Table \ref{tab:summary reflectivity after}) is 85\% for UHF1 and 80\% for UHF2, respectively, at ambient temperature~\cite{Sakaguri2022}. This transmittance can be considered a lower limit. The loss tangent is expected to decrease, and transmittance will increase at low temperatures.

\begin{figure}[tbp]
\begin{minipage}[t]{0.5\linewidth}
\centering
  \includegraphics[keepaspectratio, height=4.5cm]{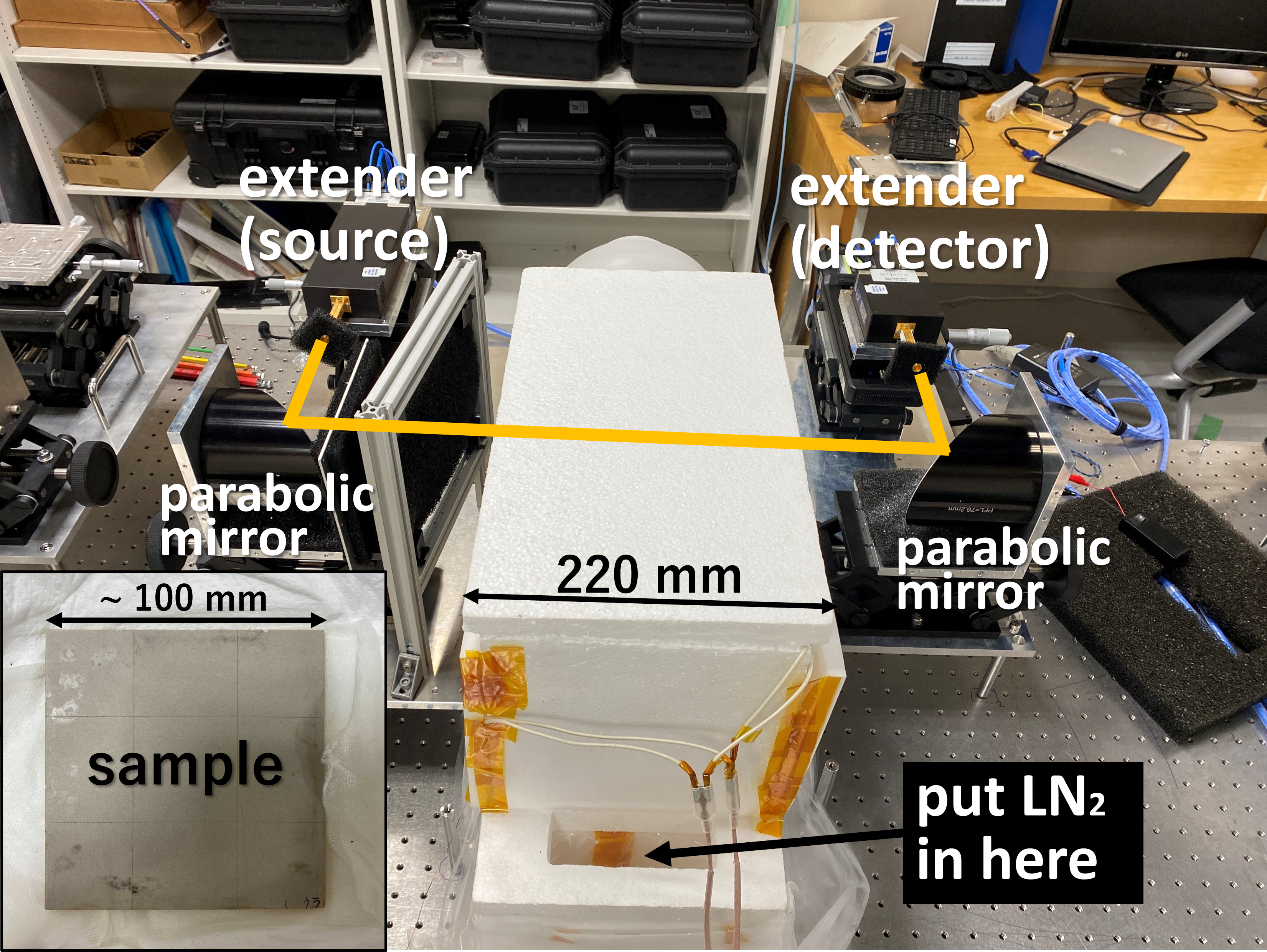}
\label{fig:LN2box photo}
\end{minipage}
\begin{minipage}[t]{0.5\linewidth}
\centering
  \includegraphics[keepaspectratio, height=4.5cm]{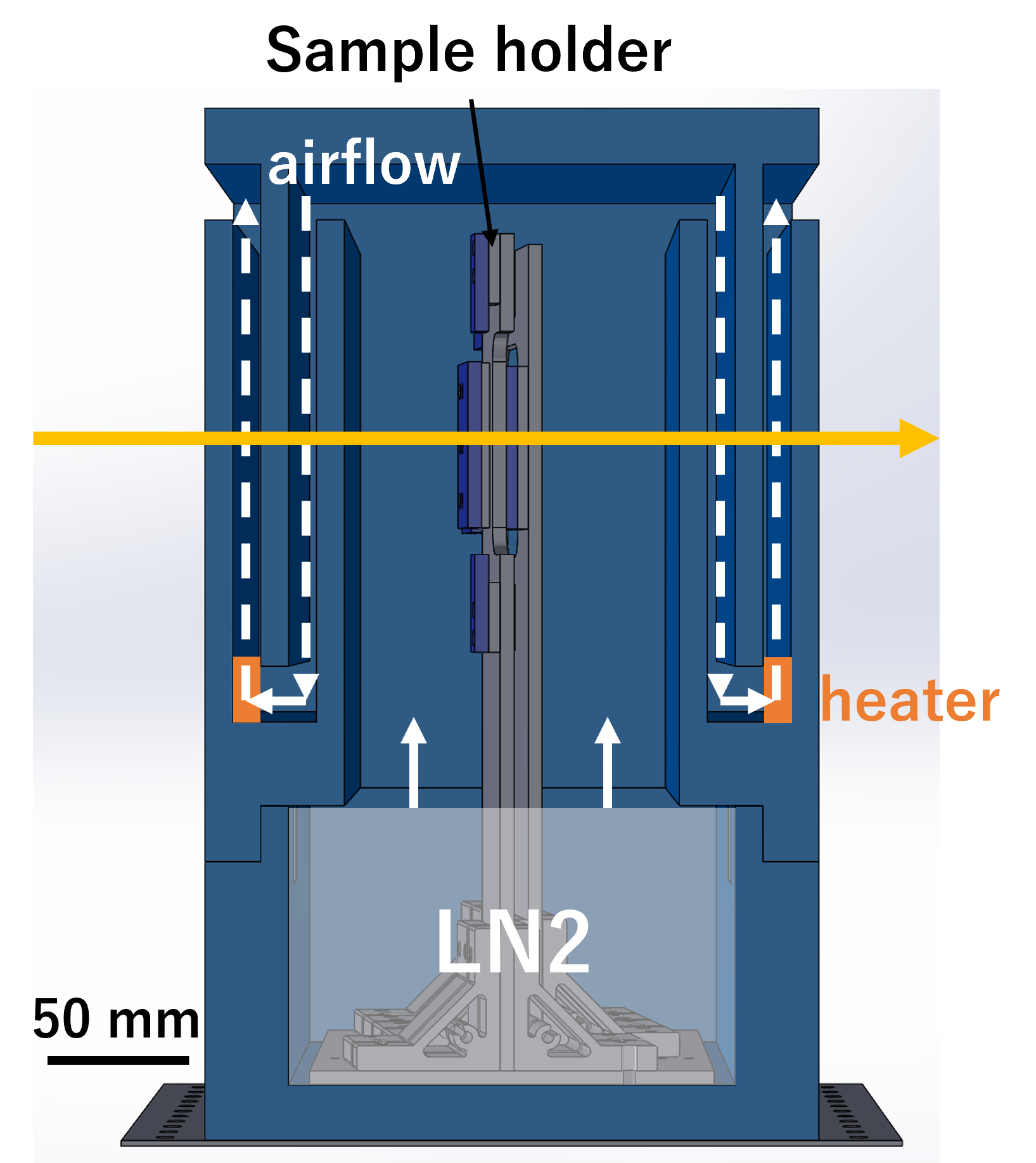}
\label{fig:LN2box schematic}
\end{minipage}
\caption{Low-temperature measurement system. Left: Photograph of the setup. The yellow arrow shows the optical path. A box with a structure made of styrofoam is placed in the optical path. Right: Schematic of the box cross-section. Oxygen-free Copper holds the sample. The width of a jagged structure is \SI{10}{mm} each. Nitrogen vapor is gradually warmed up with the long path and out.}
\label{fig:LN2box setup}
\end{figure}

We developed a setup to measure transmittance at cryogenic temperatures.
The transmittance of a UHF sample is measured with the same VNA as the room temperature measurement. It is possible to take measurements quickly since they do not need a cryostat.
There are two conditions for the measurement setup.
\begin{itemize}
    \item $\mathrm{H}_2\mathrm{O}$ condensation must not form in the optical path.
    \item The measured sample must be cooled to about \SI{100}{K}, below which we do not expect much difference in the loss tangent of coatings between \SI{100}{K} and 4, 40, or \SI{80}{\kelvin}.
\end{itemize}
Figure~\ref{fig:LN2box setup} shows the measurement setup. A box with Zotefoam (Plastazote HD30~\cite{Zotefoams}) is transparent to millimeter waves and nitrogen vapor cools the sample.
There are two factors that prevent $\mathrm{H}_2\mathrm{O}$ condensation in the optical path. The first is that the flowing nitrogen vapor is saturated, preventing the air from taking in more water inside the box. The second is that the paths from the nitrogen vapor to the outside (airflow in Figure~\ref{fig:LN2box setup}) are folded, and silicone rubber heaters SBH2117~\cite{Hakko} are placed in the middle of the paths to warm up the nitrogen vapor. This structure gives the nitrogen vapor enough time to warm up, ensuring that the box surface remains warm enough not to cool the outside air.
To cool the sample sufficiently, we hold it with oxygen-free copper, which has high thermal conductivity. Copper surrounds the four sides of the sample, and the sample is cooled by thermal contact.
To measure the transmittance accurately, two holders are prepared to side by side, one for holding the sample and the other for normalization without a sample. The transmission is first normalized by measuring a sample-free holder; then, the measurement point is shifted to the sample to measure transmittance. It eliminates the effects of copper edges, such as scattering.
The validity of the measurement system and the check for water droplets are confirmed by measuring a \SI{4}{mm} alumina slab at 90--140\,GHz.

\begin{figure}[tbp]
    \centering
    \includegraphics[width=8 cm, clip]{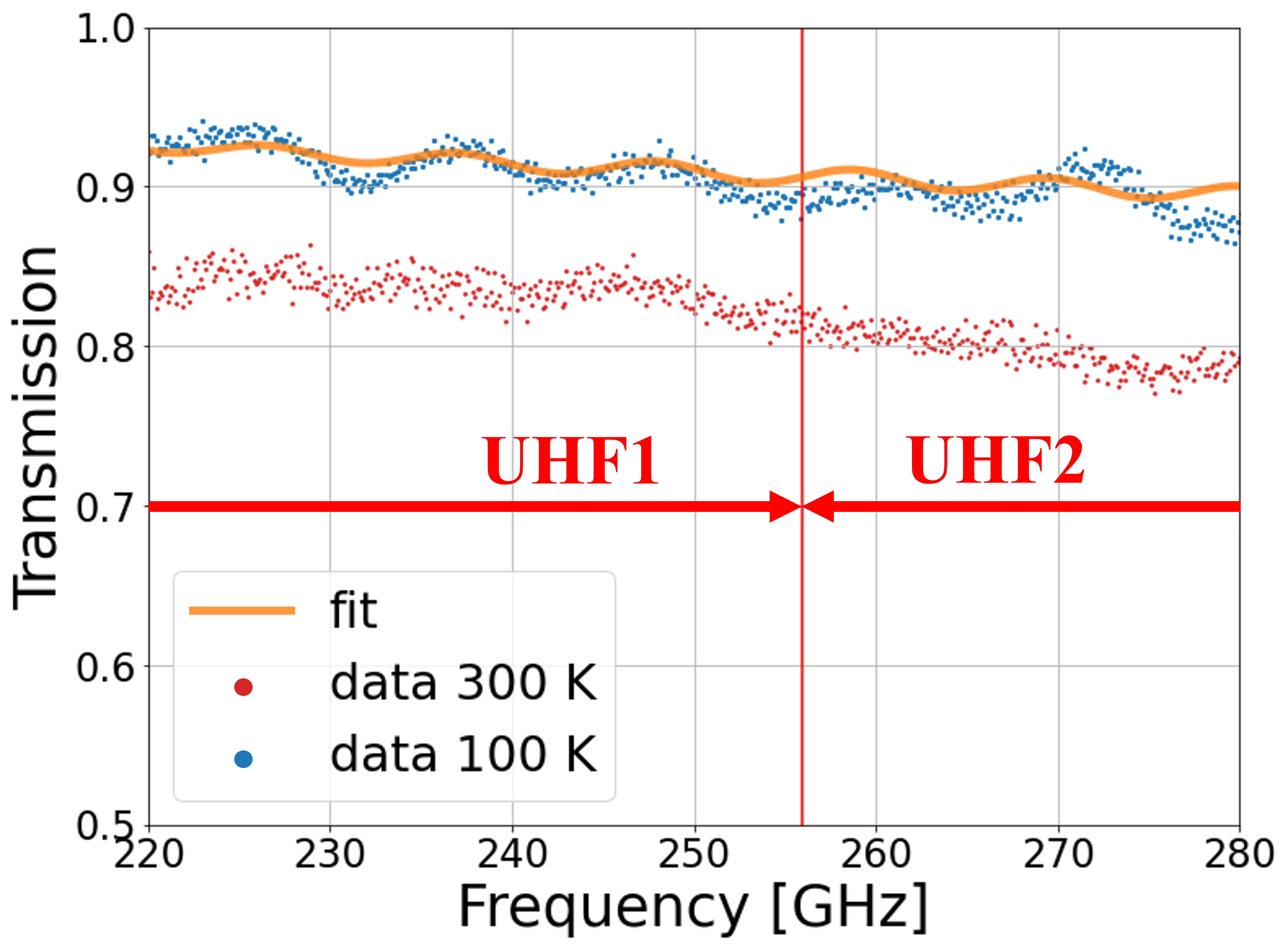}
    \caption{Transmission of S2 sample comparing measurements at \SI{300}{\kelvin} and \SI{100}{\kelvin}. The red dotted line shows the \SI{300}{\kelvin} data, and the blue dotted line shows the \SI{100}{\kelvin} data. The orange line shows the fitted value of the \SI{100}{\kelvin} data. Error is estimated from the alumina slab measurement.}
    \label{fig:trans lowtemp}
\end{figure}

Figure \ref{fig:trans lowtemp} shows the results of transmittance measurements comparing \SI{300}{\kelvin} and \SI{100}{\kelvin} behavior. It indicates that the transmittance increases at low temperatures, as expected. The average transmittance is calculated between the bands. The result of the cold measurement is transmittance of 91\% for UHF1 and 88\% for UHF2, respectively. We can estimate the absorption by calculating $1- (transmittance) - (reflectance)$, assuming that the loss tangent is independent of UHF frequency. In this measurement, the alumina thickness is \SI{4}{mm}, much thicker than the mullite and Duroid thicknesses (Table~\ref{tab:AR properties}). Therefore, it is inferred that about half of the absorption comes from the alumina by estimating an alumina loss based on the reference paper~\cite{Inoue:16}. Absorption by the AR coating for two surfaces is estimated to be $\sim 5\%$. That is sufficient for the IR filter and the HWP. The measurement itself does not distinguish between absorption and random scattering. The microspheres in the Duroid may cause the scatter, but the wavelength comparable to the sphere size is far outside the band. Also, the scattering of the Duroid taper is estimated to be negligible. We leave that to a future study. We consider the performance of large coatings to be the same since the loss tangent of each layer remains stable.

\section{Conclusion}
\label{coclusion}
We developed a two-layer AR coating for a sapphire HWP and an IR filter for CMB polarimetry at 90/150 and 220/280\,GHz frequencies. The production process is established for cryogenic use and large diameters.
We establish a quality control method. The coatings do not delaminate after a cooling cycle. The IR camera test enables quality assurance of the fabricated coatings.
Our coatings with mullite and Duroid 5880LZ reduce the reflectivity to about 2.6\%(0.9\%). The transmittance is increased to about 90\% in actual use, as evaluated at cryogenic temperature.
Such a production process for coating large optical surfaces is essential for modern CMB experiments. The HWPs and IR filters with our coating are used in the Simons Array and the Simons Observatory experiments. They are installed in the receiver at the site. This technique will also be applied to optical elements for future CMB experiments, such as CMB-S4.

\section*{Acknowledgements}
We thank Oliver Jeong for the coating inspection and advice.
This research was supported by JSR Fellowship, the University of Tokyo, FoPM, WINGS Program, the University of Tokyo, and IGPEES, WINGS Program, the University of Tokyo, JSPS Core-to-Core program grant number JPJSCCA20200003, and JSPS KAKENHI Grant Numbers 18H05539, 18KK0083, 19H00674, 23H00107, and 23H01202. We acknowledge the support by World Premier International Research Center Initiative (WPI), MEXT, Japan, International Research Center Formation Program to Accelerate Okayama University Reform (RECTOR), and the Center of
Innovation Program funded by the Japan Science and Technology Agency (JST) grant number JPMJCE1313. This work was partly supported by the Simons Foundation (Award \#457687, B.K.). Work at LBNL is supported by the U.S. Department of Energy, Office of Science, Office of High Energy Physics under contract No. DE-AC0205CH11231.

\bibliography{reference}

\end{document}